\documentclass[preprint]{aastex631}

\usepackage{amsmath}
\usepackage{subfigure}

\begin{document}

\title{Orbital decay of double white dwarfs: beyond gravitational wave radiation effects}

\author[0000-0001-8718-6925]{G. A. Carvalho}
\affiliation{Departamento de F\'isica, Universidade Tecnol\'ogica Federal do Paran\'a, Medianeira, PR, Brazil}
\affiliation{Instituto de Pesquisa e Desenvolvimento (IP\&D), Universidade do Vale do Para\'iba, 12244--000, S\~ao Jos\'e dos Campos, SP, Brazil}
\affiliation{Instituto Tecnol\'ogico de Aeron\'autica, 12228--900, S\~ao Jos\'e dos Campos, SP, Brazil}

\author{R. C. dos Anjos}
\affiliation{Departamento de Engenharias e Exatas, Universidade Federal do Paran\'a (UFPR), Pioneiro, 2153, 85950--000 Palotina}
\affiliation{PPGFISA, Universidade Federal da Fronteira Latino-Americana, PR, Brazil}
\affiliation{PPGFA, Universidade Tecnol\'ogica Federal do Paran\'a, PR, Brazil}

\author{J. G. Coelho}
\affiliation{N\'ucleo de Astrof\'isica e Cosmologia (Cosmo-Ufes) \& Departamento de F\'isica, Universidade Federal do Esp\'irito Santo, 29075--910, Vit\'oria, ES, Brazil}
\affiliation{Divis\~ao de Astrof\'isica, Instituto Nacional de Pesquisas Espaciais, Avenida dos Astronautas 1758, 12227--010, S\~ao Jos\'e dos Campos, SP, Brazil}

\author{R. V. Lobato}
\affiliation{Departamento de F\'isica, Universidad de los Andes, Bogot\'a, Colombia}
\affiliation{Department of Physics and Astronomy, Texas A\&M University-Commerce, Commerce, TX 75429, USA}

\author{M. Malheiro}
\affiliation{Instituto Tecnol\'ogico de Aeron\'autica, 12228--900, S\~ao Jos\'e dos Campos, SP, Brazil}

\author{R. M. Marinho}
\affiliation{Instituto Tecnol\'ogico de Aeron\'autica, 12228--900, S\~ao Jos\'e dos Campos, SP, Brazil}

\author{J. F. Rodriguez}
\affiliation{Escuela de F\'isica, Universidad Industrial de Santander, Ciudad Universitaria, Bucaramanga, Colombia}
\affiliation{ICRANet, Piazza della Repubblica 10, I-65122 Pescara, Italy}

\author{J. A. Rueda}
\affiliation{ICRANet, Piazza della Repubblica 10, I-65122 Pescara, Italy}
\affiliation{ICRA, Dipartimento di Fisica, Sapienza Universit\`a  di Roma, Piazzale Aldo Moro 5, I-00185 Rome, Italy}
\affiliation{ICRANet-Ferrara, Dipartimento di Fisica e Scienze della Terra, Universit\`a degli Studi di Ferrara, Via Saragat 1, I-44122 Ferrara, Italy}
\affiliation{Dip. di Fisica e Scienze della Terra, Universit\`a degli Studi di Ferrara, Via Saragat 1, I-44122 Ferrara, Italy}
\affiliation{INAF, Istituto di Astrofisica e Planetologia Spaziali, Via Fosso del Cavaliere 100, I-00133 Rome, Italy}

\author{R. Ruffini}
\affiliation{ICRANet, Piazza della Repubblica 10, I-65122 Pescara, Italy}
\affiliation{ICRA, Dipartimento di Fisica, Sapienza Universit\`a  di Roma, Piazzale Aldo Moro 5, I-00185 Rome, Italy}
\affiliation{INAF, Viale del Parco Mellini 84, I-00136 Rome, Italy}



\begin{abstract}

The traditional description of the orbital evolution of compact-object binaries, like double white dwarfs (DWDs), assumes that the system is driven only by gravitational wave (GW) radiation. However, the high magnetic fields with intensities of up to gigagauss measured in WDs alert a potential role of the electromagnetic (EM) emission in the evolution of DWDs. We evaluate the orbital dynamics of DWDs under the effects of GW radiation, tidal synchronization, and EM emission by a unipolar inductor generated by the magnetic primary and the relative motion of the non-magnetic secondary. We show that the EM emission can affect the orbital dynamics for magnetic fields larger than megagauss. We applied the model to two known DWDs, SDSS J0651+2844 and ZTF J1539+5027, for which the GW radiation alone does not fully account for the measured orbital decay rate. We obtain upper limits to the primary's magnetic field strength, over which the EM emission causes an orbital decay faster than observed. The contribution of tidal locking and the EM emission is comparable, and together they can contribute up to $20\%$ to the measured orbital decay rate. We show that the gravitational waveform for a DWD modeled as purely driven by GWs and including tidal interactions and EM emission can have large relative dephasing detectable in the mHz regime of frequencies relevant for space-based detectors like LISA. Therefore, including physics besides GW radiation in the waveform templates is essential to calibrate the GW detectors using known sources, e.g., ZTF J1539+5027, and to infer binary parameters.

\end{abstract}

\keywords{(stars:) white dwarfs --- (stars:) binaries (including multiple): close --- stars: magnetic field --- gravitational waves}


\section{Introduction}\label{sec:int}

Our galaxy hosts a predicted number of $(1$--$3)\times 10^8$ double white dwarfs (hereafter DWDs) \cite{Nelemans2001,Nelemans2005,Maoz2012May}, of which observational facilities have detected only about $100$. This situation can improve thanks to forthcoming space-based detectors of gravitational waves (GWs) like the Laser Interferometer Space Antenna (LISA), which expects to detect the GW radiation driving the dynamics of compact, detached DWDs  (see, e.g., Refs. \cite{Stroeer2006Sep,Korol2022Apr}). The detection and analysis of GW signals need the development of gravitational waveform templates that accurately encode the physics driving the binary dynamics. The traditional description of the orbital evolution of compact-object binaries, like DWDs, assumes that the gravitational wave (GW) radiation of two point-like masses orbiting the common center of mass is an accurate description of the binary dynamics, neglecting any other interactions. However, the orbital evolution is affected by additional effects like the dark matter background (see, e.g., \cite{2015PhRvD..92l3530P,2017PhRvD..96f3001G}) and the electromagnetic (EM) emission (see, e.g., \cite{Marsh2005Oct,2018ApJ...868...19W}). We focus in this article on the effects of the latter.

There is mounting observational evidence that the components of DWDs can be highly magnetized. Depending on the binary component masses, the merger of a DWD may not lead to a prompt type Ia supernova (SN) but a newborn, massive, fast rotating, highly magnetic WD (see, e.g., \cite{2018ApJ...857..134B}). Mergers of DWDs have been proposed as progenitors of ZTF J190132.9+145808.7 \cite{2021Natur.595...39C} and the recently discovered isolated, highly magnetic, rapidly rotating WD (rotation period of $70.32$ s), SDSS J221141.80+113604.4 (see \cite{2021ApJ...923L...6K} for details). These rotation rates are consistent with the theoretical predictions for DWD merger remnants, in agreement with the many works published in the last decade about the theory of highly magnetic, massive, and fast WDs from DWD mergers \cite{2012PASJ...64...56M,2012IJMPS..18...96C,2013ApJ...772L..24R,2014PASJ...66...14C,2014ApJ...794...86C,doi:10.1142/S021827181641025X,2016JCAP...05..007M,2017MNRAS.465.4434C,2017A&A...599A..87C,2018ApJ...857..134B,Otoniel2019Jul,2020MNRAS.492.5949S,2020MNRAS.498.4426S,2020ApJ...895...26B}.

The above extreme properties of some WDs have led to the proposal that DWD mergers can power low-energy gamma-ray bursts (GRBs). The prompt gamma-ray emission arises from the transient activity of the magnetosphere during the merger, the infrared/optical transient from the merger ejecta, and an extended X-ray and radio emission powered by the WD central merger remnant \cite{2019JCAP...03..044R}. In addition, high-energy neutrinos may be the product of cosmic-ray acceleration in DWD mergers and newborn pulsars \cite{2016ApJ...832...20X}. The rapid rotation and strong magnetic fields can accelerate particles to energies higher than petaelectronvolt (PeV; i.e., $10^{15}$ eV), and the surrounding material can naturally generate ultrahigh-energy cosmic rays (UHECR) with energies larger than exaelectronvolt (EeV; i.e., $10^{18}$ eV), in particular, with a heavy composition \cite{2016ApJ...826...97P,2021JCAP...10..023D}. The rotational magnetic instability surrounding the source can lead to the formation of hot, magnetized corona and high-velocity outflows. Additionally, the low volume of the surrounding material facilitates the escape of UHECRs from the environment \cite{2016ApJ...826...97P, 2013ApJ...773..136J, 2014MNRAS.438..169B, 2020PhRvD.102l3013V}. The operation of the near generation of multi-messenger observatories like the Cherenkov Telescope Array (CTA) \cite{2011ExA....32..193A}, POEMMA \cite{2017ICRC...35..542O}, and IceCube \cite{2011arXiv1111.2736T} will shed more light on several high-energy scenarios and interpretations for understanding particle acceleration in a DWD merger.

Given all the above, in this article, we analyze the dynamics of DWDs in the pre-merger stage under the action of GW emission, tidal interactions, and electromagnetic (EM) emission. The inclusion of a large variety of possible emissions besides the GW radiation could complicate the analysis of the results and hide the essential physics we would like to spot here. Therefore, we emphasize here only the effects of the EM emission on the binary dynamics using the \textit{unipolar inductor model} (UIM) \cite{1969ApJ...156...59G} applied to DWDs (see, e.g., \cite{Wu2002,Osso2006,Lai2012}). The EM emission in the UIM originates from the energy dissipation of the closed circuit formed by the magnetized primary star, the non-magnetic secondary, and the magnetic field lines. The motion of the secondary relative to the magnetic field lines of the primary generates the electromotive force (EMF) that drives the current through the magnetic field lines (see, e.g., \cite{Wu2002,Lai2012}). We refer the reader to \cite{Lai2012} (and references therein) for estimates of the EM emission from the UIM in a variety of compact-object binaries.

We show with specific examples that the EM emission by the UI overcomes the emission from a hot WD and magnetic-dipole braking. Such an EM emission is comparable to the quadrupolar GW radiation by two orbiting point-like masses. Therefore, we include the EM emission in the binary dynamics and quantify its contribution to the rate of orbital decay. We show that the EM emission can significantly affect the binary dynamics, accounting for a sizable part of the orbital decay measured in some compact DWDs and the GW properties (e.g., phase, intensity). Therefore, it is of paramount relevance to understand and model the physical phenomena that drive the binary dynamics to develop astrophysical waveform templates useful to detect and infer binary parameters from GW signals (see, e.g., \cite{2022arXiv220103226B}).

We organize the article as follows. In Section~\ref{sec:2}, we recall the aspects of the UIM that are relevant for the modeling of the DWD dynamics, estimate the EM dissipation for fiducial values of the masses and magnetic field, solve (numerically) the equations of motion, and compare with the orbital decay of a pure GW-radiation-driven dynamics. Section~\ref{sec:3} analyzes within the UIM two known DWDs, SDSS J0651+2844 and ZTF J1539+5027. We analyze the constraints on the system given by the measured orbital decay, obtain upper limits to the primary's magnetic field, and estimate the contribution of tidal synchronization and EM emission to the orbital decay. We quantify in Section~\ref{sec:4} the effect of the EM emission in the phase evolution of the GWs. Finally, we present in Section~\ref{sec:5} the conclusions of this article.

\section{Unipolar inductor and orbital dynamics}\label{sec:2}

We follow the general framework of the UIM presented in \cite{Wu2002} and use the associated EM dissipation estimated in \cite{Lai2012}. The binary system is composed of a magnetic primary with mass $M_1$, radius $R_1$, and magnetic moment $\mu_1$, and a non-magnetic secondary with mass $M_2$ and radius $R_2$. Unless otherwise stated, we estimate the WD radius from the mass-radius relation presented in \cite{Carvalho2018b,carvalho_2019}. The secondary is synchronous, so it has angular velocity $\omega_s = \omega_0$, where
\begin{equation}\label{eq:omega0}
    \omega_0 = \sqrt{\frac{G M}{r^3}},
\end{equation}
is the orbital angular velocity according to Kepler's third law. The primary is asynchronous with angular velocity $\Omega$ measured by the parameter $\alpha = \Omega/\omega_0$. Hereafter, we denote with $M = M_1+M_2$ and $r$ the binary's total mass and orbital separation.

The evolution of the binary system under the combined (non-linearly coupled) GW radiation, tides, and EM emission losses in the UIM is obtained from energy and angular momentum conservation which lead to the system of equations \cite{Wu2002}
\begin{eqnarray}
    \frac{\dot{\omega}_0}{\omega_0}&=&-\frac{\dot{P}}{P} = \frac{1}{g(\omega_0)}\left[\dot{E}_{\rm GW}-\frac{L}{1-\alpha}\right],\label{eq10}\\
    \frac{\dot{\alpha}}{\alpha}&=& -\frac{1}{g(\omega_0)}\left\{\dot{E}_{\rm GW}-\frac{L}{1-\alpha}\left[1+\frac{g(\omega_0)}{\alpha I_1\omega_0^2}\right]\right\},\label{eq11}
\end{eqnarray}
where $P=2\pi/\omega_0$ is the orbital period, $L$ is the EM power released by the circuit, and $\dot{E}_{\rm GW}$ is the rate of energy loss via GW radiation for a system of two point-like masses in circular orbit
\begin{align}\label{eq:Lgw}
    \dot{E}_{\rm GW} &= -\frac{32}{5} \frac{G}{c^5} \left(\frac{q}{1+q}\right)^2 M_1^2 r^4 \omega_0^6 \nonumber\\  
   &= -\frac{32}{5} \frac{G}{c} \left(\frac{q}{1+q}\right)^2 M_1^2 \left(\frac{G M \omega_0}{c^3}\right)^{4/3},
\end{align}
where we have used Eq. (\ref{eq:omega0}) in the second equality, and
\begin{equation}\label{gfunc}
    g(\omega_0) = -\frac{1}{3}\left(\frac{q^3}{1+q}G^2 M_1^5\omega_0^2\right)^{1/3}\left[1-\frac{6}{5}(1+q)\left(\frac{R_2}{r}\right)^2\right],
\end{equation}
with $q=M_2/M_1$ the binary's mass-ratio.

The above model of the binary dynamics remains valid to the point when either Roche lobe overflow of the secondary or merger takes place. Therefore, the maximum orbital angular velocity of the system is
\begin{equation}\label{eq:omegamax}
    \omega_0^{\rm max}=\sqrt{\frac{G M}{r_{\rm min}^3}},
\end{equation}
being $r_{\rm min} = {\rm Max}(r_L,r_{\rm mrg})$, where according to Eggleton's formula for the Roche lobe \cite{1983ApJ...268..368E}
\begin{equation}\label{eq:roche}
    r_L = \frac{0.6 q^{2/3} + \ln(1+q^{1/3})}{0.49 q^{2/3}}R_2,
\end{equation}
and $r_{\rm mrg}=R_1+R_2$. For instance, for a $0.6+0.6 M_\odot$ binary, with $R_1=R_2 \approx 7.8 \times 10^8$ cm, $r_L \approx 2.06 \times 10^9$ cm, and $r_{\rm mrg} \approx 1.56 \times 10^9$ cm. For these figures, Eq. (\ref{eq:omegamax}) leads to $\omega_0^{\rm max}\approx 0.13$ rad s$^{-1}$, corresponding to a minimum orbital period of $46.43$ s. In all the examples presented in this article, the orbital dynamics is analyzed far from any of the above two physical situations.

The equations of motion (\ref{eq10})--(\ref{eq11}) account for the torques due to the EM emission and from tides (see \cite{Wu2002} for details). We now recall the EM power of the UIM. The motion of the conductive secondary into the primary's rotating magnetosphere induces an electromotive force ${\cal E} = 2 R_2 |\vec{E}|$, where the electric field and associated electric potential $U$ through the secondary star are
\begin{equation}
    \vec{E} =\frac{\vec{v}\times \vec{B}}{c},\quad U=2 R_2 |E|,
\end{equation}
being
\begin{equation}\label{eq:vrel}
    \vec{v}=r(\omega_0-\Omega)\hat{\phi}=(G M \omega_0)^{1/3}(1-\alpha)\hat{\phi},
\end{equation}
and we have used Eq. (\ref{eq:omega0}) in the second equality. The total energy dissipation is \cite{Wu2002}
\begin{equation}\label{EMlum}
L=2I^2\mathcal{R},
\end{equation}
where the factor $2$ accounts for the upper and lower parts of the circuit, $\mathcal{R}$ is the total resistance of the system, and $I=U/\mathcal{R}$ is the electric current.

\begin{figure}
  \centering
  \includegraphics[width=0.8\hsize,clip]{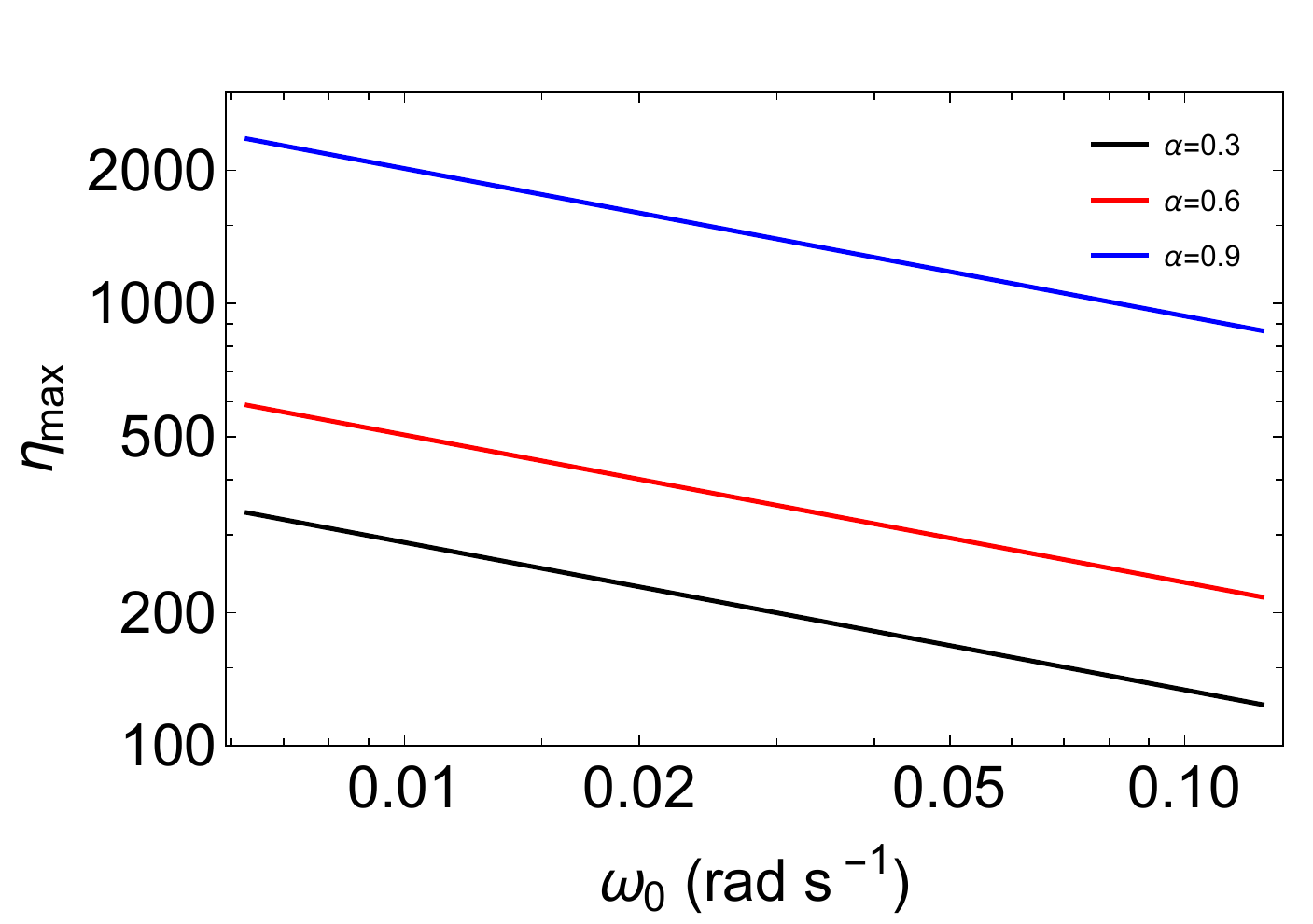}
  \caption{Value of $\eta_{\rm max}$ as a function of $\omega_0$, given by Eq. (\ref{eq:etamax}), for selected values of $\alpha$.}\label{fig:etamax}
\end{figure}

\citet{Lai2012} has shown that a high twist of the magnetic field causes the disruption of the magnetic flux tubes, hence short-circuiting the system. The azimuthal twist is given by $\xi_\phi=-B_{\phi+}/B_z=16v/(c^2 \mathcal{R})$, where $B_{\phi+}$ is the toroidal magnetic field generated by the current in the circuit on the upper side of the primary. Therefore, we limit the twist parameter to $\xi_\phi \lesssim 1$ (i.e., $\mathcal{R} \gtrsim 16 v/c^2$), so that the circuit remains active. Bearing the above in mind, we parametrize the resistance in terms of the value given by the impedance of free space, i.e., 
\begin{equation}\label{eq:Rtot}
    {\cal R} = \frac{4\pi}{c} \frac{1}{\eta},
\end{equation}
which leads to
\begin{equation}\label{eq:xi}
    \xi_\phi = \eta\frac{4}{\pi}\frac{v}{c} = \frac{4}{\pi}\left(\frac{G M \omega_0}{c^3}\right)^{1/3} (1-\alpha)\eta,
\end{equation}
where we have used Eq. (\ref{eq:vrel}) to obtain the second equality. We limit the value of $\eta$ so to have $\xi \leq 1$ during the entire evolution. Therefore, $\eta_{\rm max}$ is
\begin{equation}\label{eq:etamax}
    \eta_{\rm max} = \frac{\pi}{4}\left(\frac{c^3}{G M \omega_0}\right)^{1/3} \frac{1}{1-\alpha}.
\end{equation}
Figure \ref{fig:etamax} shows the value of $\eta_{\rm max}$ as a function of the $\omega_0$, for selected values of $\alpha$.

Having set all the above, the EM power (\ref{EMlum}) derived in \cite{Lai2012} can be written as
\begin{equation}\label{Lai}
    \frac{L}{1-\alpha}=\frac{2}{\pi c} (1-\alpha) \eta\, \omega_0^2 \frac{\mu_1^2 R_2^2}{r^4}.
\end{equation}
Normalizing the physical quantities in Eq. \eqref{Lai} to fiducial parameters for DWDs, the EM power reads
\begin{equation}\label{WDemission}
    \frac{L}{1-\alpha}=7.72\times 10^{32}\,\left(\frac{\tilde{B}}{10^{6}\,{\rm G}}\right)^2 \left(\frac{R_1}{10^9~{\rm cm}}\right)^6 \times \left(\frac{R_2}{10^9~{\rm cm}}\right)^2 \left(\frac{M_\odot}{M}\right)^{4/3}\left(\frac{100\, {\rm s}}{P}\right)^{14/3} {\rm erg ~s^{-1}},
\end{equation}
where we have used the primary's magnetic moment $\mu_1=B R_1^3$, with $B$ the magnetic field, and have introduced
\begin{equation}\label{eq:Btilde}
    \tilde{B} \equiv \sqrt{(1-\alpha)\eta}B,
\end{equation}
a quantity that encloses the degeneracy among $\alpha$, $\eta$, and $B$ in the Eqs. (\ref{eq10}) and (\ref{eq11}).

\begin{figure}
  \centering
  \includegraphics[width=0.8\hsize,clip]{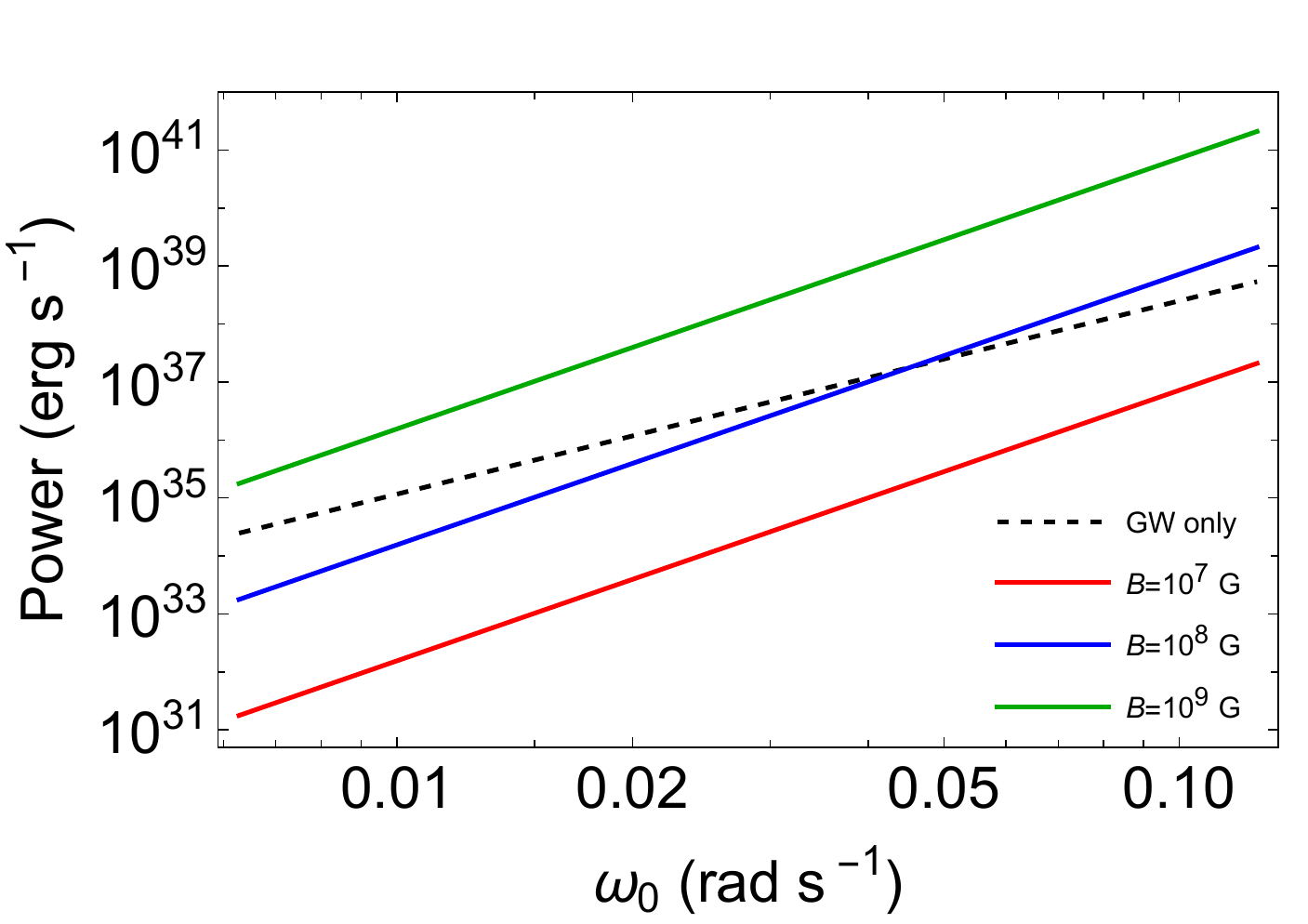}
  \caption{EM power as a function of orbital angular frequency given by Eq. (\ref{WDemission}), for selected values of the primary's magnetic field. In this example, we set $\eta = 100$, and the binary is mass-symmetric with $M_1=M_2=0.6M_\odot$. The mass-radius relation is taken from \cite{Carvalho2018b}. For comparison, we also show the GW power (black dashed curve) given by Eq. (\ref{eq:Lgw}).}\label{luminosity}
\end{figure}

Figure \ref{luminosity} shows the EM power (\ref{WDemission}) as a function of the orbital angular velocity, in the case of $\alpha=0.9$, and $M_1=M_2=0.6 M_\odot$ ($R_1 = R_2 = 7.79\times 10^8$ cm), for selected values of the magnetic field strength ranging from $10^6$ G to $10^9$ G. For instance, for a magnetic field $B=10^9$ G, $\eta = 100$, and orbital period of $50$ s and $300$ s, Eq. (\ref{WDemission}) leads, respectively, to an EM power of $2.66\times 10^{39}$ erg s$^{-1}$ and $6.23\times 10^{35}$ erg s$^{-1}$. This luminosity is much larger than the blackbody luminosity of a hot WD with surface temperature of $10^4$ K, $L_{\rm BB} = 4 \pi R_1^2 \sigma T^4 \approx 4.33\times 10^{30}$ erg s$^{-1}$, or the EM emission owing to magnetic dipole braking, respectively, $L_{\rm dip} \sim (1/c^3)B^2 R_1^6 \Omega^4 \approx 1.36\times 10^{36}$ erg s$^{-1}$ and $1.05\times 10^{33}$ erg s$^{-1}$.

\begin{figure*}
  \centering
  \includegraphics[width=0.49\hsize,clip]{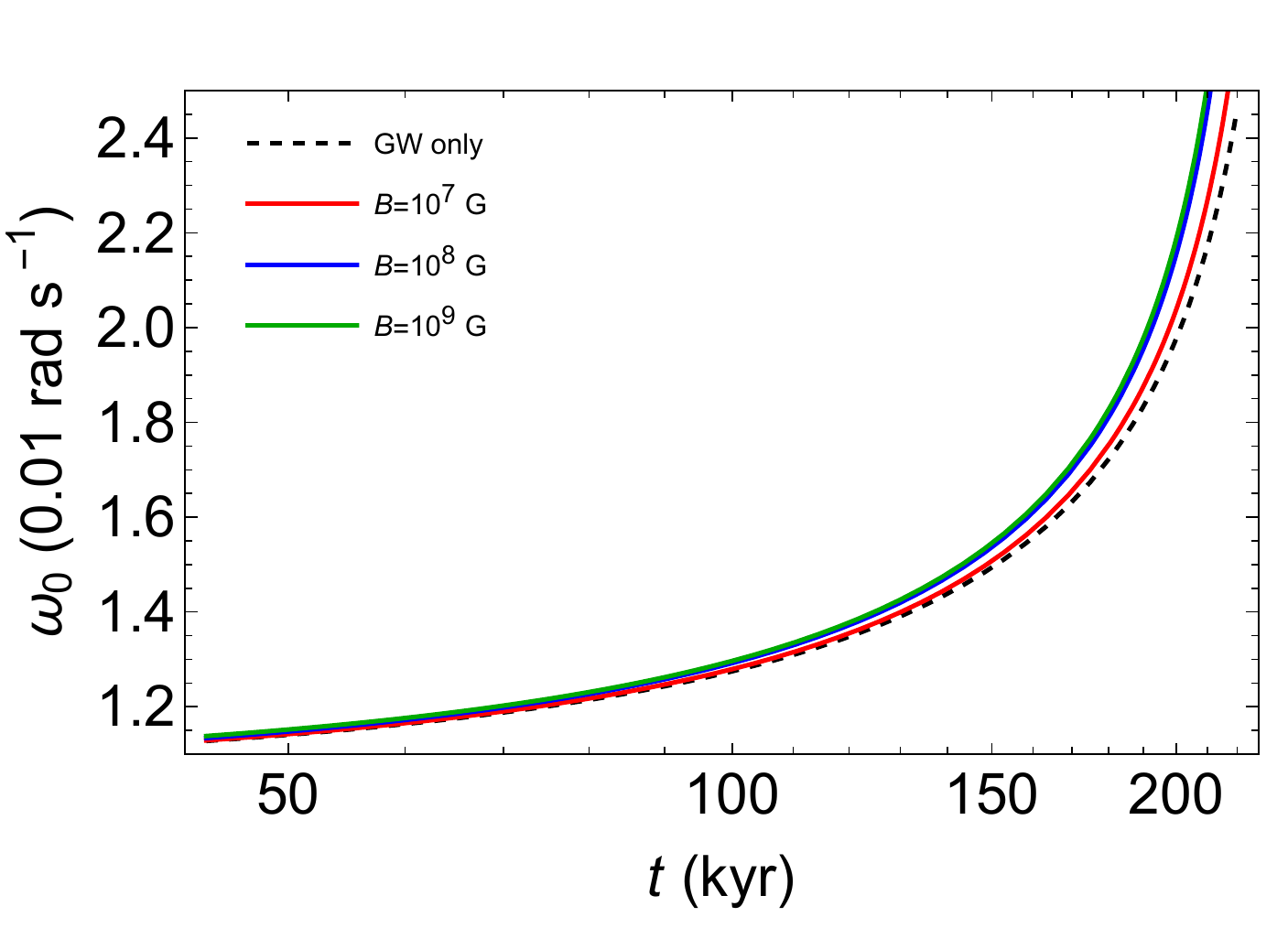}\includegraphics[width=0.49\hsize,clip]{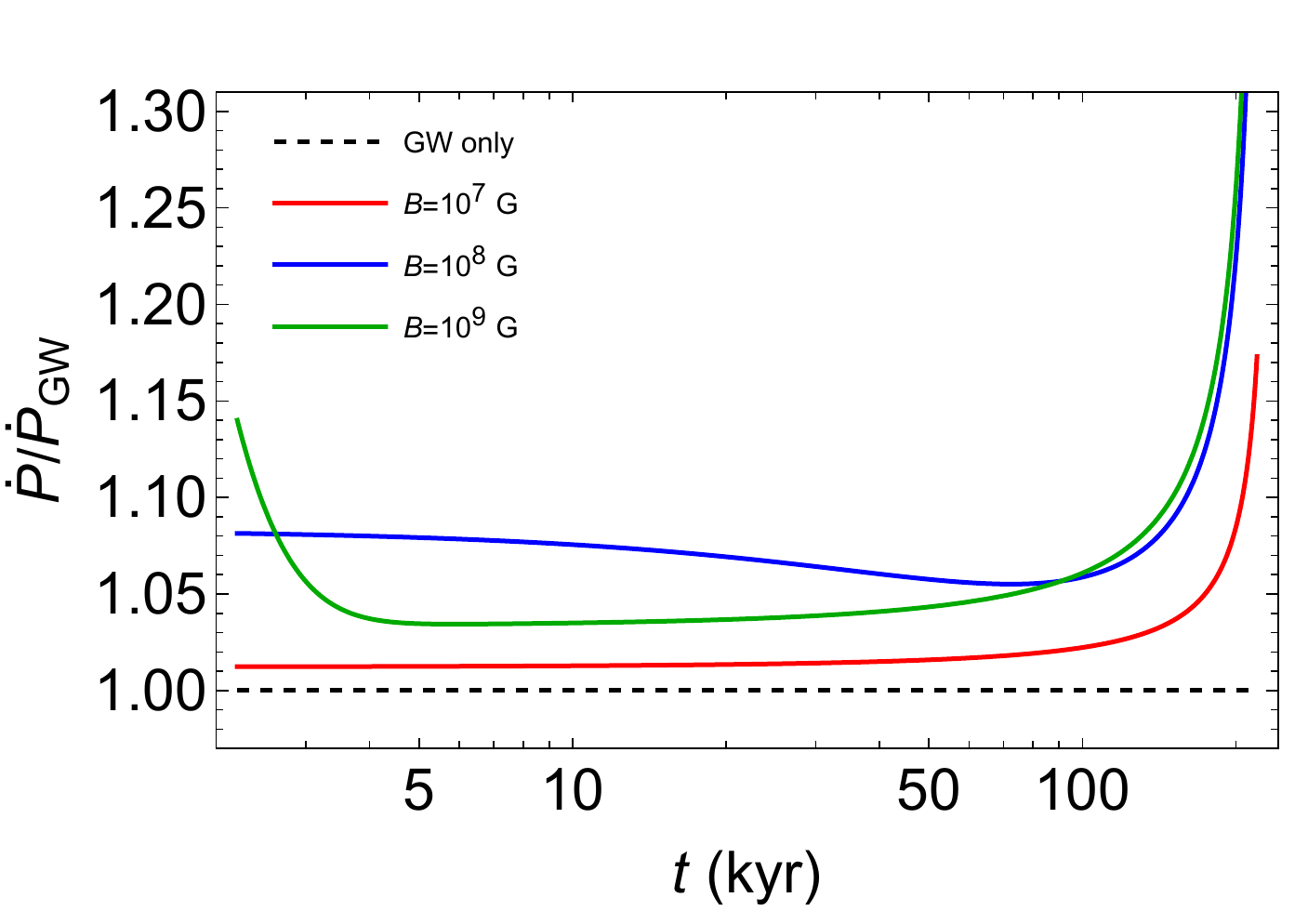}
  \includegraphics[width=0.49\hsize,clip]{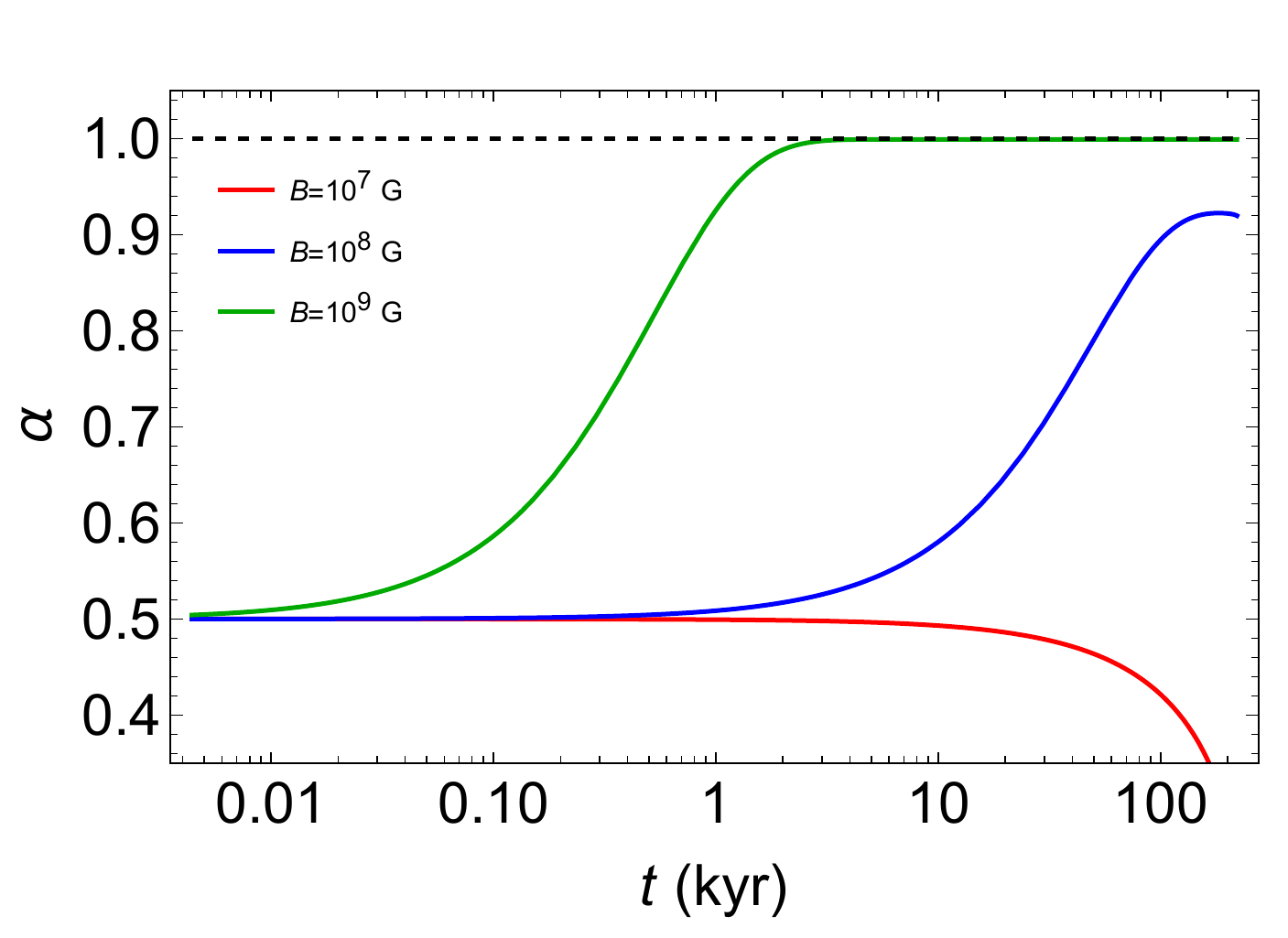}
  \includegraphics[width=0.49\hsize,clip]{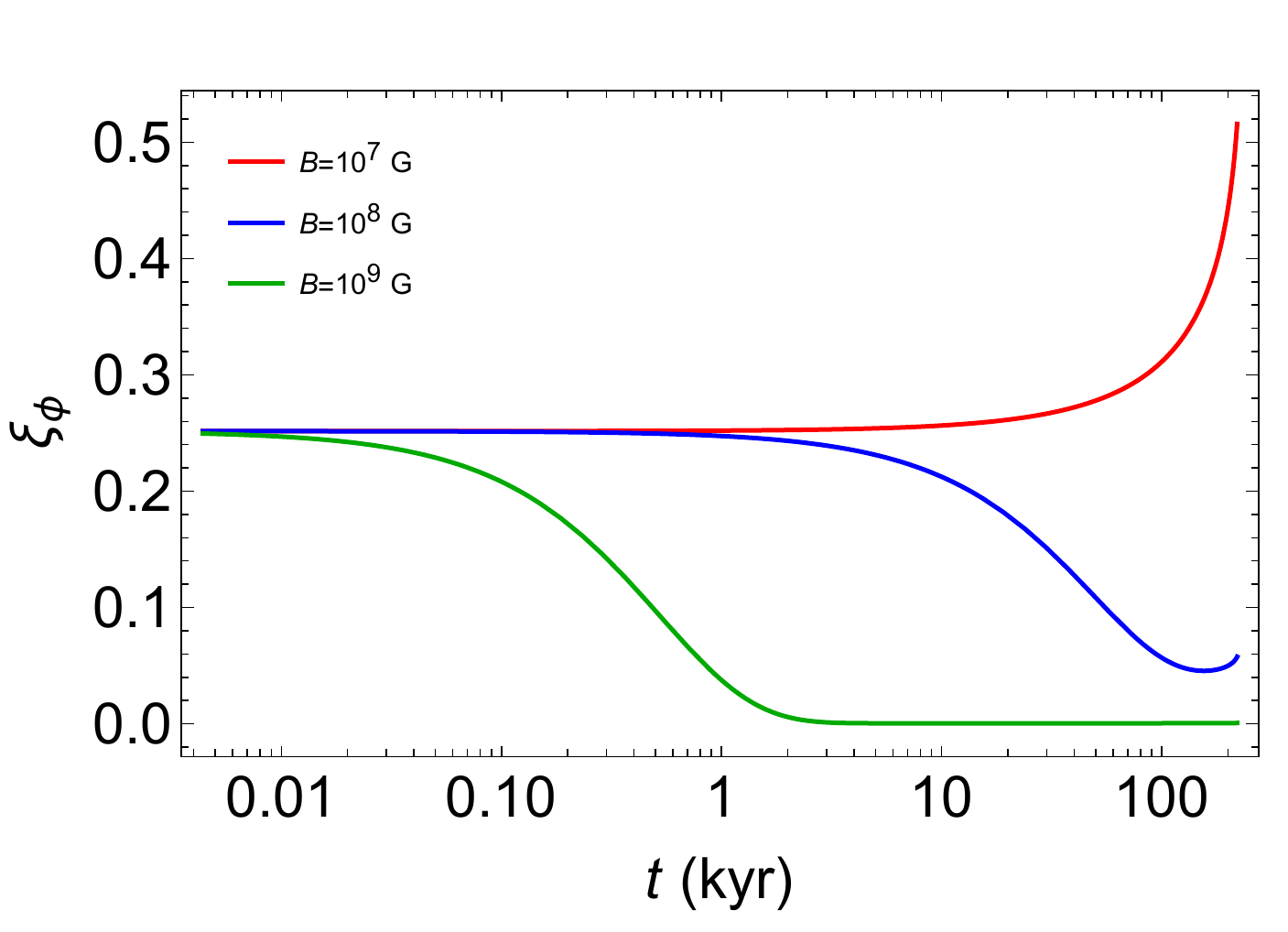}
  \caption{Examples of numerical solution of the equations of motion (\ref{eq10})--(\ref{eq11}), for selected values of the primary's magnetic field. In these examples, we set $\eta = 100$, the binary is mass-symmetric with $M_1=M_2=0.6M_\odot$, and assign an initial ($t=0$) values for the orbital period and the degree of synchronization of the primary, respectively, $P(0) = 10$ min (i.e., $\omega_0(0) = 0.0105$) and $\alpha(0) = 0.5$. The mass-radius relation is taken from \cite{Carvalho2018b}. For comparison, we also show the solution of the equations of motion when only GW radiation is considered, i.e., the solution to Eq. (\ref{eq:PdotGW}). Upper left: orbital evolution, $\omega_0$. Upper right: orbital decay rate, $\dot{P}$, normalized to the value for the case of only GW radiation, $\dot{P}_{\rm GW}$. Lower left: evolution of the primary's synchronization, $\alpha$. Lower right: evolution of the twist parameter, $\xi_\phi$.}\label{OE}
\end{figure*}

Figure \ref{luminosity} also indicates that for magnetic fields of the order of $10^{9}$ G, the EM emission of the UIM can even overcome the GW emission before merger, so it largely affects the orbital dynamics at those evolution stages. For lower magnetic fields, the EM emission lowers but remains comparable to the GW emission at high frequencies, i.e., for compact binaries. Under these conditions, the orbital evolution is not driven only by GW radiation, but rather by a coupling between GW and EM emission. Figure \ref{OE} shows the evolution of $\omega_0$ with time, for an initial orbital period of $10$ minutes. We compare the results of the orbital dynamics given by Eq. \eqref{eq10}, which accounts for GWs, tides and EM emission, with the case when the dynamics is purely driven by GW radiation. In the latter case, the rate of orbital decay is given by
\begin{equation}\label{eq:PdotGW}
 \dot{P}_{\rm obs} = \dot{P}_{\rm GW} = -\frac{96}{5}(2\pi)^{8/3}\frac{G^{5/3}}{c^5}\frac{M_1 M_2}{M^{1/3}} P^{-5/3}.
\end{equation}
In this specific example, for magnetic fields $\gtrsim 10^6$ G, the tidal locking and the EM emission starts to affect the orbital dynamics, and for fields $\gtrsim 10^8$ G the effects become noticeably large. 

\section{Constraining the magnetic field in observed double white dwarfs}\label{sec:3}

\begin{table*}
    \caption{\label{tab1} Example of DWDs with short orbital periods that are targets for LISA-like missions. An upper limit for the magnetic field of the UIM can be set if the DWD has measured $P$, $\dot{P}$, $M_1$ and $M_2$. See main text for details.}
    \hspace{-2cm}\begin{tabular}{l|c|c|c|c|c}
      Binary & $M_1/M_\odot$ & $M_2/M_\odot$ & $P$ (s) & $\dot{P}_{\rm obs}$ (s s$^{-1}$) & Refs.\\
        \hline
       ZTF J1539+5027 & $0.610^{+0.017}_{-0.022}$ & $0.210^{+0.014}_{-0.015}$ & $414.79 \pm 2.9\times 10^{-6}$ & $(-2.373 \pm 0.005)\times 10^{-11}$ & [1] \\
       \hline
       SDSS J0651+2844 & $0.50\pm 0.04$ & $0.26\pm 0.04$ & $765.2 \pm 5.5\times 10^{-5}$ & $(-9.8 \pm 2.8)\times 10^{-12}$ & [2] \\
       \hline
    \end{tabular}
    \tablecomments{[1] \cite{Burdge2019Jul}, [2] \cite{Brown2011Jul,Hermes2012Sep}}
\end{table*}

Since the orbital evolution of the binary is affected by the EM emission and tides, it is theoretically possible to infer the magnetic field or at least to put constraints on it from measurements of the orbital decay rate. Therefore, given measurements of the orbital period, $P$, the spin-down rate of the orbital period, $\dot{P}_{\rm obs}$, and the binary component masses, $M_1$ and $M_2$ (the corresponding WD radii are assumed to be known from the mass-radius relation), we can constrain the magnetic field. For this task, we request that the spin-down rate predicted by the UIM, which includes the effect of the GWs, the EM emission and tides, does not exceed the measured orbital decay, $\dot{P}_{\rm obs}$, i.e.,
\begin{equation}\label{eq:Pdotconstraint}
    \dot{P}_{\rm obs} = \dot{P},
\end{equation}
where $\dot{P}$ is the period decay given by the model, which is obtained from the solution of the system of equations \eqref{eq10}--\eqref{eq11}. In this light, we analyze two known compact DWDs, ZTF J1539+5027 \cite{Burdge2019Jul} and SDSS J0651+2844 \cite{Brown2011Jul,Hermes2012Sep}.

\subsection{SDSS J0651+2844}\label{sec:3A}

\begin{figure*}
    \centering
    \includegraphics[width=\hsize,clip]{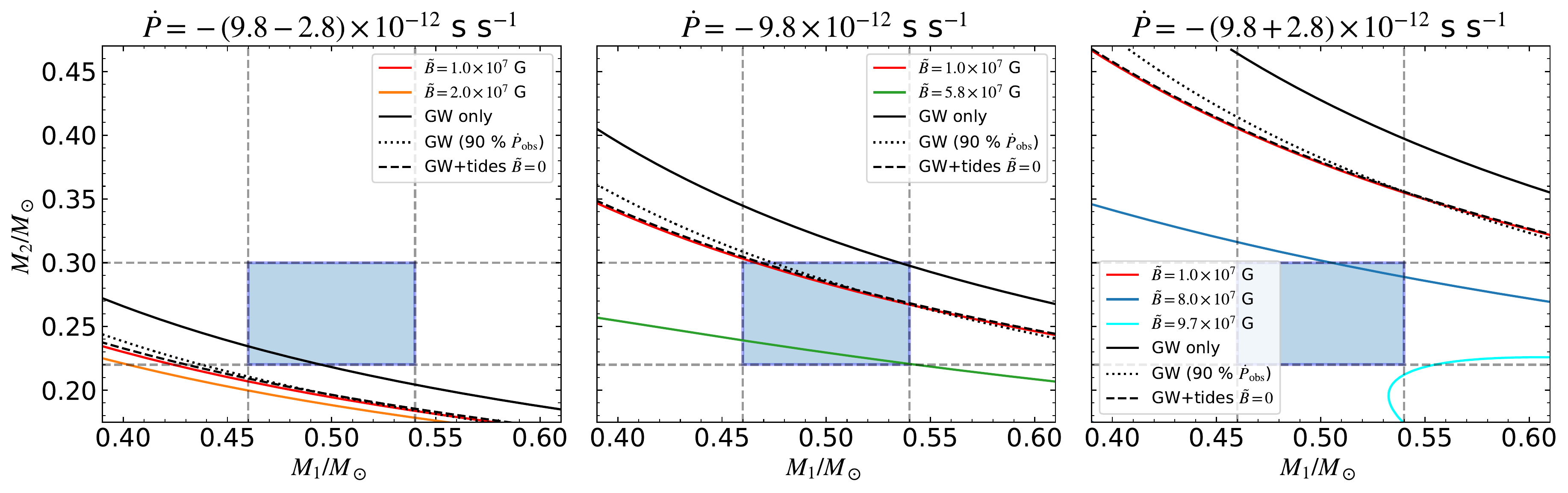}
    \caption{Constraints on the binary masses and magnetic field for SDSS J0651+2844. We have used the values inferred from photometric and spectroscopic measurements of the orbital period, decay rate, and masses reported in \cite{Hermes2012Sep} (see also Table \ref{tab1}). The blue rectangle represents the $1\sigma$ uncertainties on the masses. Left: constraints considering the lower limit for the decay rate, $\dot{P}=-(9.8-2.8)\times 10^{-12}$~s~s$^{-1}$, which gives the lower limit to the contribution of $\tilde{B}$ and tides. Pure GW radiation is consistent with $1\sigma$ errors in the masses. Center: constraints considering the central value of the decay rate, $\dot{P}=-9.8 \times 10^{-12}$~s~s$^{-1}$. Right: constraints considering the upper limit for the decay rate, $\dot{P}=-(9.8 + 2.8)\times 10^{-12}$~s~s$^{-1}$, which we use to estimate the upper limit on $\tilde{B}$. Summarizing, the current decay rate is consistent with a pure GW-driven dynamics, but the uncertainties on the mass measurements and $\dot{P}$ are broad enough to allow solutions of GW-emission admixed with a UI and tides, although the absence of Zeeman splitting in the spectra rule out magnetic field strengths $B>10^6$ G.}
    \label{fig:m2vsm1DWD1}
\end{figure*}

Table \ref{tab1} lists the parameters $M_1$, $M_2$, $P$ and $\dot{P}_{\rm obs}$ of SDSS J0651+2844, reported in \cite{Brown2011Jul, Hermes2012Sep}.
Given values of $\dot{P}$ and $P$, Eq. \eqref{eq:Pdotconstraint} with $\dot{P}$ given by Eq. \eqref{eq10}, gives a relation between $M_1$ and $M_2$ for every given value of $\tilde{B}$.  Figure \ref{fig:m2vsm1DWD1} shows examples of the constraints on the masses obtained from the orbital period and decay rate of SDSS J0651+2844. We compare the results of the UIM with the case of pure GW radiation, i.e., when using Eq. \eqref{eq:PdotGW}, the case with $90\%$ of GW radiation and the case with GW radiation plus tides, i.e., Eqs. \eqref{eq10} and \eqref{eq11} with $L=0$. The agreement with the observational data requires that the $M_2$-$M_1$ relations cross the measurements of $M_1$ and $M_2$ represented within $1\sigma$ error by the blue rectangle. The pure GW-driven evolution is consistent with the data, but the current statistical uncertainties in the measured masses and $\dot{P}$ allow alternative explanations of the binary dynamics including additional physics to the GW emission, like UI and tides, for a relatively wide parameter space. Therefore, tighter constraints on $\dot{P}$ are needed to conclude more on the sole basis of timing. The absence of Zeeman splitting in the spectra of J0651+2844 rule out magnetic fields $B>10^6$ G.

In \cite{Hermes2012Sep, Burdge2019Jul}, it has been pointed out that, indeed, a sizable portion of the observed orbital decay might arise in these DWDs from tidal interactions. Besides GWs and tides, the model studied in this work takes also into account the effect of EM emission from an active UI in the binary. Figure \ref{fig:m2vsm1DWD1} shows three curves of the UIM, and the case of including GWs and a full tidal locking but without EM emission ($\tilde{B}=0$). We recall that as the synchronization parameter $\alpha$ changes with time (see, e.g., Fig. \ref{OE}), the value of $\tilde{B}$ must be considered as a constraint at the observational period. For $\tilde{B}=10^7$ G (red curve), the effect of the EM emission is relatively small, so the dynamics is dominated by GW radiation and tidal synchronization. This model nearly follows the curve of the model $\dot{P}_{\rm GW} = 0.9 \dot{P}_{\rm obs}$, which suggests that roughly $90\%$ of the orbital decay is due to GW radiation, and the remaining $10\%$ to tidal locking. For $\tilde{B}=5.8\times 10^7$ G (green curve), the EM emission has considerable effects in the dynamics, as shown by the difference of this curve {in comparison with the examples with lower magnetic field values}. In fact, the data do not favor models with high values of {$\tilde{B}$ as shown by the upper limit on $\tilde{B}$ set by the $3\sigma$ upper limit on $\dot{P}$. For $\tilde{B}\gtrsim 9.7\times 10^7$~G, the $M_1$-$M_2$ curve for those cases lie outside the rectangle of $1\sigma$ uncertainties in the masses}. Although due to the non-linearity of the model is not possible to separate the individual contributions to the $\dot{P}$, we have checked that a curve in which $77\%$ of the orbital decay is due to GW radiation approaches the green curve in the lower right part of the rectangle {(middle panel)}, suggesting that the contribution of GW radiation in the green-curve model could be around that value, and the remaining $\approx 23\%$ is shared in comparable amounts by the tidal interactions and EM emission.

\subsection{ZTF J1539+5027}\label{sec:3B}

Table \ref{tab1} lists the parameters $M_1$, $M_2$, $P$ and $\dot{P}_{\rm obs}$ of ZTF J1539+5027, reported in \cite{Burdge2019Jul}. In this case, the masses of the DWD components are not known from photometric and/or spectroscopic measurements. The reported values of the masses have been obtained in \cite{Burdge2019Jul} from crossed-information by the measured spectroscopic radial-velocity semi-amplitudes, the constraint to the mass-radius relation of the primary combined with the ratio of the primary's radius to the semi-major axis, $R_1/r$ inferred from lightcurve modelling, and constraints imposed by the binary chirp mass assuming that the orbital decay is $100\%$ driven by GW radiation (solid black curve), or $90\%$ (dotted black curve) assuming a $10\%$ from tidal interaction considering full synchronization of both the primary and the secondary.

\begin{figure}
    \centering
    \includegraphics[width=\hsize,clip]{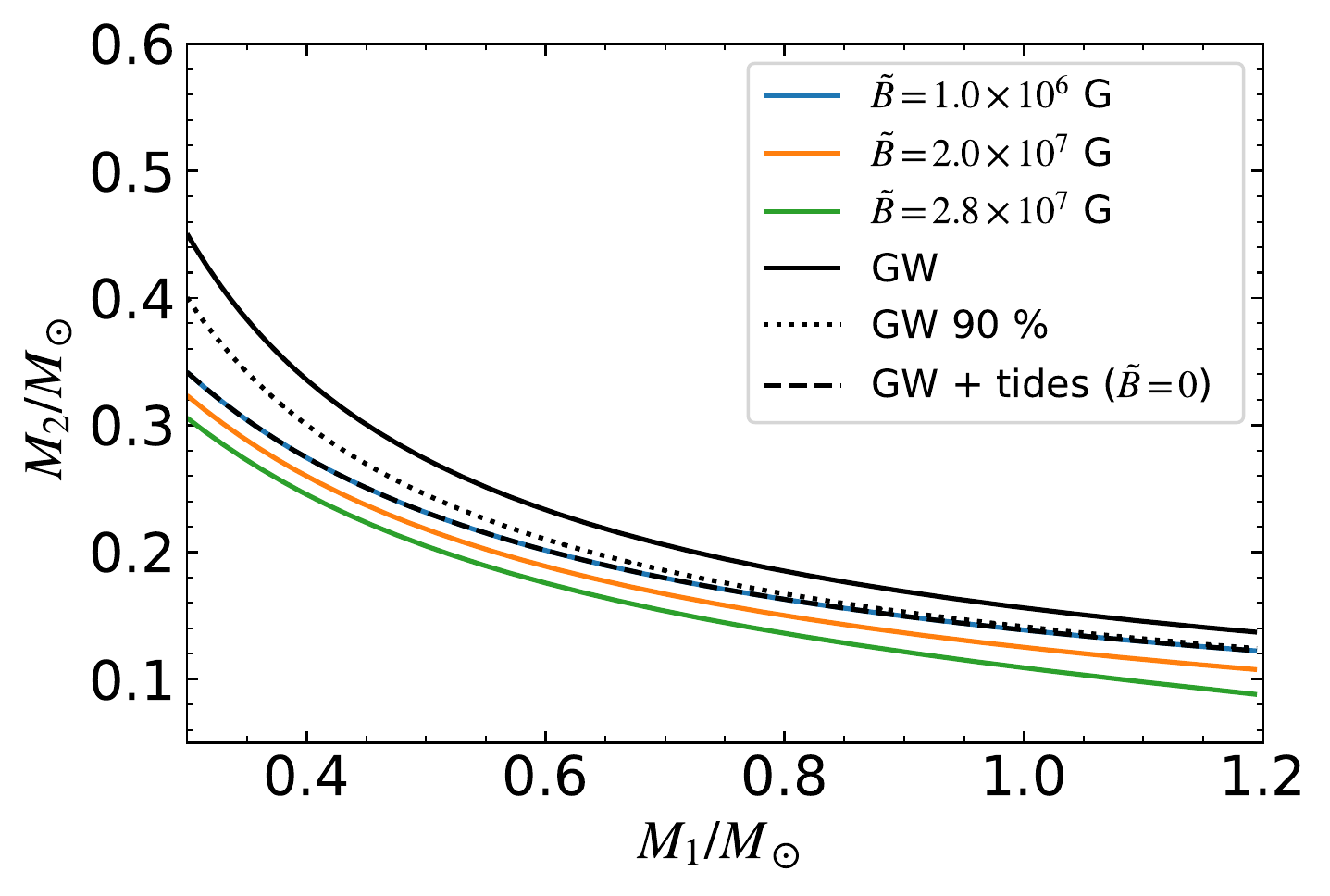}
    \caption{Constraint on the binary masses and magnetic field of the primary in ZTF J1539+5027. We have used the orbital period and decay rate reported in \cite{Burdge2019Jul} and listed in Table \ref{tab1}.}
    \label{fig:m2vsm1DWD2}
\end{figure}

Since in this case the mass values depend on the adopted model, we apply the present model considering GW radiation, tides, and the EM emission by the UI, and cross-check it with the other independent constraints. We plot in Fig. \ref{fig:m2vsm1DWD2} the results for $\tilde{B}=1.0\times 10^6$ G (blue curve), $2.0\times 10^7$ G (orange curve), and $2.8\times 10^7$ G (green curve). In doing this, we adopted in the function $g(\omega_0)$ given by Eq. \eqref{gfunc}, the observational constraint on the secondary's radius, $R_2/r = 0.28$, as reported in \cite{Burdge2019Jul}. For $\tilde{B}\lesssim 10^7$ G, the EM emission effect is relatively small. In fact, the blue curve partly overlaps with the black dotted curve $\dot{P}_{\rm GW} = 0.9 \dot{P}$, with the remaining $\approx 10\%$ dominated by the partial tidal synchronization. For larger values of $\tilde{B}$, the EM emission has appreciable effects. Indeed, models with $\tilde{B}\gtrsim 2.8\times 10^7$ G are not favored by the observational data, since the resulting $M_1$-$M_2$ relation falls below the lower limit imposed by the 50\% contour level of the mass-radius constraint shown in \cite{Burdge2019Jul}. Figure \ref{fig:m2vsm1DWD2} shows that within the above range of allowed values of $\tilde{B}$, some solutions allow slightly lower values for the masses with respect to the solution considered in \cite{Burdge2019Jul} of nearly $90\%$ of $\dot{P}$ arising from GWs and $10\%$ from tidal synchronization.

\section{Intrinsic time-domain phase evolution of gravitational waves}\label{sec:4}

Having shown that physics besides GW radiation, e.g., tidal and EM emission, can have appreciable effects on the orbital dynamics, we analyze in this section the effect that this could have on the gravitational waveform.

The evolution of the orbital angular frequency is quite slow for a considerable part of the lifetime of the binary. Consequently, these systems can be considered as quasi-monochromatic GW emitters. It is worth mentioning that if the source is exactly monochromatic (given the sensitivity of the detector) the nature of the system cannot be determined by observing its gravitational radiation. We will consider only the evolution stages when the system is not monochromatic, that is, only those orbital frequency regimes of the system in which an interferometer is capable of detecting changes in frequency.

For a quasi-monochromatic source, the intrinsic parameters of the gravitational waveform template are the frequency, $f$, its time derivative, $\dot{f}$, and the wave amplitude, $h_0$ \cite{2002ApJ...575.1030T}. The amplitude depends both on intrinsic parameters (e.g., the binary mass) and also on external parameters like the distance to the source. The first two parameters ($f$ and $\dot{f}$) define the intrinsic time-domain phase evolution of the GWs as \cite{2013PhRvD..87h4035D}
\begin{equation}\label{eq:Q}
    Q_{\omega} = \frac{\omega^2}{\dot{\omega}}= -\frac{2\pi}{\dot{P}} =2\pi \frac{d N}{d\ln \omega},
\end{equation}
which provides information on the rate change of the GW phase per logarithmic change in frequency. Here, $\omega=2\omega_0$ is the GW angular frequency. The quantity $Q_{\omega}$ is useful to compare the phase evolution of two waveforms given it is invariant under phase and time shifts. The integral of the difference between the value of $Q_\omega$ of two waveforms gives their relative \textit{dephasing}. For a binary emitting only GW in the pure point-like quadrupole approximation, the phase evolution $Q_\omega^{\rm GW}$ is
\begin{equation}\label{eq:Qgw}
       Q_{\omega}^{\rm GW} = \frac{5}{3\nu} 2^{-\tfrac{7}{3}} \biggl(\frac{G M \omega}{c^3}\biggr)^{-\tfrac{5}{3}} = \frac{5}{48\nu}\biggl(\frac{GM\omega_0}{c^3}\biggr)^{-\tfrac{5}{3}},
\end{equation}
where $\nu = M_1 M_2/M^2 = q/(q+1)^2$ is the so-called symmetric mass-ratio. For example, a binary with $M=1.2$~$M_\odot$, $q=1$ ($\nu = 1/4$), driven only by GW emission, has $Q_\omega^{\rm GW}= 3.2\times10^{12}$ at $f = \omega/(2\pi) = 1$~mHz.

As already mentioned, the frequency evolution of a binary under GW, tidal interaction and EM emission is different from a pure GW-radiation-driven dynamics. Therefore, the GW phase evolution is also different. The slower a system changes its frequency, the more cycles it achieves before changing its frequency, i.e., $Q_\omega$ is larger. Since the evolution of the binary under pure GW emission is slower (see Fig. \ref{OE}), we can infer that $ Q_{\omega}^{\rm UIM}<Q_{\omega}^{\rm GW}$. 

\begin{figure}
  \centering
  \includegraphics[width=\hsize,clip]{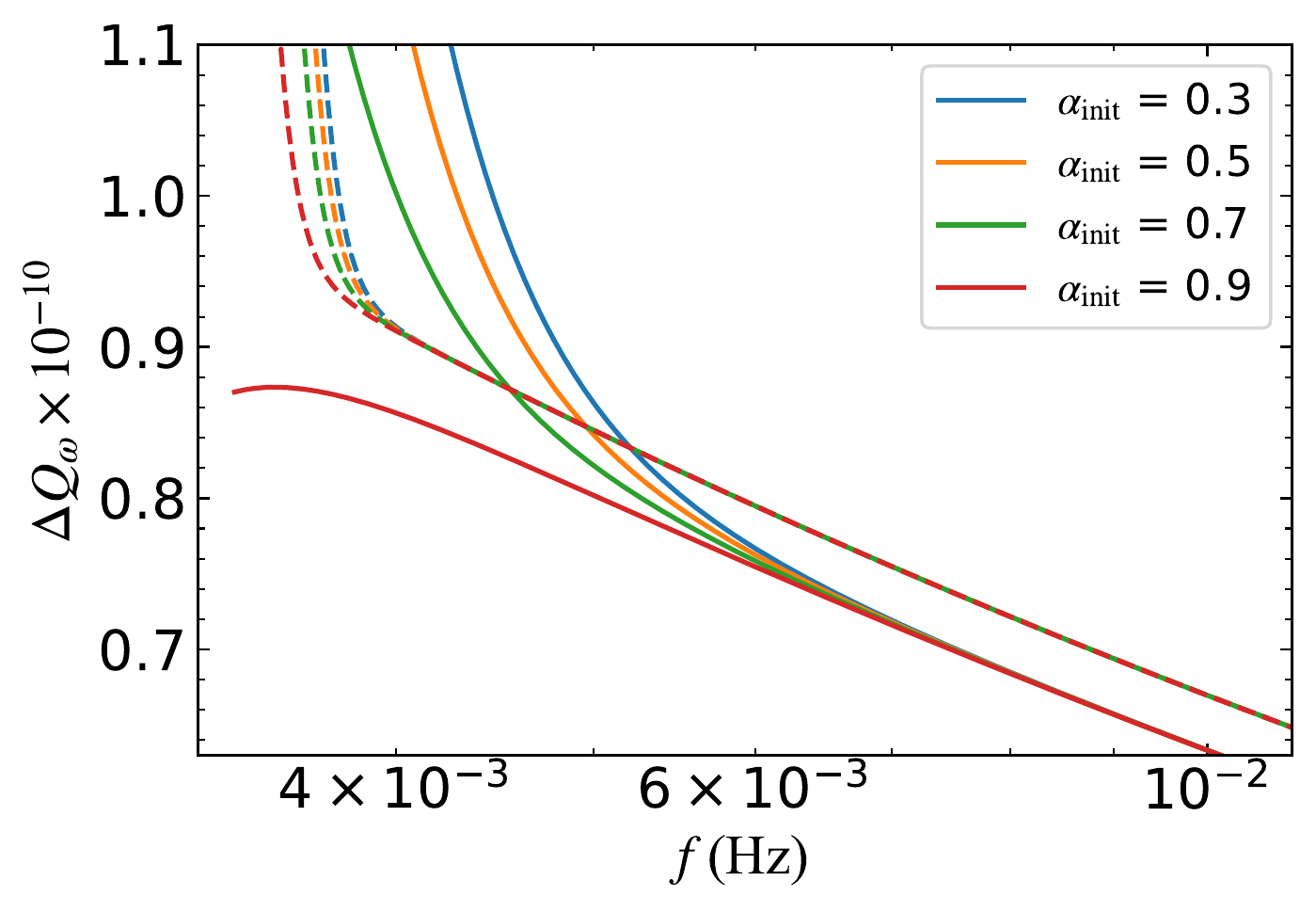}
  \caption{Difference in the intrinsic time-domain phase evolution between the two models $\Delta Q_\omega \equiv Q_\omega^{\rm GW} - Q_\omega^{\rm UIM}$, for a binary with $M=1.2$~$M_\odot$, $q=1$, $\eta=100$, and two selected values of the magnetic field $B = 8\times10^7$~G (solid curves), $B = 2\times10^8$~G (dashed curves). The different colors correspond to different initial values of $\alpha$ used in the numerical integration. The intrinsic time-domain phase parameter encapsulates two of the intrinsic observable obtained from GW data. Since $\Delta Q_\omega \gg 1$, the UIM waveform gets out of phase with respect to the waveform in the case of only GW radiation. A sufficiently large dephasing can be detected by LISA for appropriate conditions (see main text for details)}\label{fig:OE_Qw}
\end{figure}

Figure \ref{fig:OE_Qw} shows the difference in the parameter $Q_\omega$ between the UIM and the pure GW emission model, $\Delta Q_\omega \equiv Q_\omega^{\rm GW} - Q_\omega^{\rm UIM}$, as a function of the GW frequency, for $M=1.2$~$M_\odot$, $q=1$, $\eta=100$, two selected values of the magnetic field, $B = 8\times10^7$~G (continuous curves) and $B = 2\times10^8$~G (dashed curves), and for different initial values of synchronization parameter $\alpha$. For each magnetic field case, the different curves corresponding to different $\alpha_{\rm init}$ converge rapidly. This is a consequence of the existence of a \textit{quasi-attractor} different from unity in the dynamics of synchronization parameter, $\alpha$ (see, e.g., Fig. \ref{OE}). Furthermore, the intrinsic time-domain evolution is affected for increasing values of the magnetic field.

The considerable difference between the models implies a relative dephasing of the gravitational waveforms, even when the changes in frequency are small. Suppose that we observe the above system at a GW frequency of $6$~mHz, i.e., at an orbital period of $P = 5.6$~min, and the synchronization is $\alpha=0.8$. After 2 years of evolution, the GW frequency has changed  $1.57\times10^{-3}$~\%, in the case of the UIM model with magnetic field $B=8\times10^7$~G, and $1.47\times10^{-3}$~\%, in the case of GW emission. For the former frequency change, the difference in phase of the waveforms is $\Delta \Phi\approx \Delta Q_\omega d\ln\omega = 1.48\times10^5$~rad. For a magnetic field of $2\times10^8$~G, the system changes its frequency $1.88\times10^{-3} $~\% in the same time interval and the dephasing between the two waveforms increases to $\approx 5.19\times10^5$~rad.

From the observational viewpoint, we can distinguish the two systems by the fact that the observable $\dot{f}$ is different at the same frequency. This difference can be measured by GW detectors like LISA \cite{2017arXiv170200786A}. The error in estimating $\dot{f}$ by using matched-filtering method is \cite{2002ApJ...575.1030T}
\begin{equation}\label{eq:deltafdot}
    \Delta \dot{f}_{\rm error} \approx 0.43 \left(\frac{10}{\langle \rho \rangle }\right) \frac{1}{T_{\rm obs}^2},
\end{equation}
where $\langle \rho \rangle$ is the signal-to-noise ratio (SNR) accumulated in the observing time, $T_{\rm obs}$. The SNR for quasi-monochromatic sources can be estimated as \cite{maggiore2008gravitational} 
\begin{equation}
    \langle \rho^2 \rangle = \frac{6}{25} \frac{\hat{h}_c^2(f_{\rm obs})}{f_{\rm obs} S_n(f_{\rm obs})},
\end{equation}
where the factor $6/25$ comes from averaging over the angles and considering two Michelson interferometers, $f_{\rm obs}$ is the observed GW frequency, $S_n(f)$ is the power spectrum density of the detector noise, and $\hat{h}_c$ is the \textit{reduced} characteristic amplitude \cite{1998PhRvD..57.4535F}
\begin{equation}
    \hat{h}_c (f) = h_0(f)\sqrt{N} = h_0(f)\sqrt{f T_{\rm obs}},
\end{equation}
with $h_0 = 4\nu (G M/d c^2)(\pi G  M f/c^3)^{2/3}$ the GW amplitude and $d$ is the distance to source.

Using the same system at a GW frequency $f=6$~mHz, the differences between the UIM and the pure GW emission model for $B=8\times10^7$~G and $B=2\times10^8$~G are, respectively, $\Delta \dot{f} = 9.26\times 10^{-17}$ Hz s$^{-1}$ and $\Delta \dot{f} = 3.91\times10^{-16}$ Hz s$^{-1}$. Suppose that the binary is located at a distance of $1$ kpc, so after $2$ years of observations by LISA, it could reach an SNR $\langle\rho\rangle\approx 246$, and Eq. \eqref{eq:deltafdot} gives $\Delta\dot{f}_{\rm error} \approx 4.38\times 10^{-18}$ Hz s$^{-1}$. These figures imply that LISA could discriminate between the two waveforms.

This effect can be used to calibrate the detectors observing known astrophysical sources, and pinpoint additional effects besides GW radiation in the orbital dynamics. For instance, we have shown that the gravitational waveform has the imprint of the EM emission, so the detection of GW radiation from these binaries can constrain the magnetic fields present. The above can be accomplished if the additional effects like the EM emission are accounted in the gravitational waveform templates.

\section{Conclusions}\label{sec:5}

We have shown in this article that both tidal locking and EM emission from the UI mechanism can contribute to binary dynamics of DWDs as much as the GW radiation. Therefore, physics besides GW radiation can cause large effects in the orbital decay rate and, consequently, on the GW waveforms. At the same time, this can modify the outcome of population synthesis models and the DWD merger delay time distribution, relevant for the massive WD population from merging DWDs and the double-degenerate channel of type Ia supernova.

Particularly relevant is the effect of the EM emission on the orbital decay for a magnetic field parameter $\tilde{B}\gtrsim 10^7$ G, leading to the possibility of constraining the magnetic field from measurements of the orbital decay in known DWDs. We applied the present model to two DWDs. For SDSS J0651+2844 (see section \ref{sec:3A} and Fig. \ref{fig:m2vsm1DWD1}), we obtain an upper limit $\tilde{B}\approx 6\times 10^7$ G, and for ZTF J1539+5027 (see section \ref{sec:3B} and Fig. \ref{fig:m2vsm1DWD2}), the upper limit is $\tilde{B}\approx 2.8\times 10^7$ G. We have estimated that in these systems tidal locking and EM emission can be of the same order and might have a combined contribution of $\sim 20\%$ to the orbital decay.

The fact that the contribution of physics beyond the GW radiation is already evident in known binaries motivated us to quantify the effect of the different orbital dynamics on the GW time-domain phase evolution, i.e., on the gravitational waveform (see Section \ref{sec:4}). We have shown that the waveforms obtained assuming two different dynamics, one driven totally by GW radiation and one driven by GW, tides and EM emission have an extremely diverse phase evolution (see Fig. \ref{fig:OE_Qw}) that can be measured by LISA. The sensitivity of LISA to distinguish differences in the phase evolution of different waveforms is particularly important for known sources, since the accurate modeling of the templates will allow an accurate test of the detector itself. For instance, as pointed out in \cite{Burdge2019Jul}, a \textit{crucial verification source} for LISA is ZTF J1539+5027, which emits GWs with frequency $f \approx 5$ Hz and could be detected with an accumulative large SNR of about $143$ in four years of LISA observations. For this SNR and observing time, Eq. \eqref{eq:deltafdot} states that the error in estimating $\dot{f}$ by matched-filtering will be $\Delta \dot{f}_{\rm error} \approx 2 \times 10^{-18}$ Hz s$^{-1}$. This value of $\Delta \dot{f}_{\rm error}$, together with our estimates in Section \ref{sec:4}, imply that LISA will be sensitive enough to discriminate between different models for this system. The difference in $\dot{f}$ at the GW frequency of this source between a model accounting for GW radiation, tidal interactions and EM emission and a model with only GW radiation is in the range $10^{-17}$--$10^{-15}$ Hz s$^{-1}$ for magnetic fields $10^7$--$10^9$ G. In addition, the well-constrained binary inclination angle constrains the relative amplitude of two GW polarizations, and the measured orbital decay already constraints the chirp mass \cite{Burdge2019Jul} and, as shown in this article (see Fig. \ref{fig:m2vsm1DWD1}), physics beyond GW radiation.

There are additional targets of interest for potential studies of this topic, e.g., the eclipsing DWD ZTF J2243+5242, with an orbital period of $8.8$ min and masses $M_1 = 0.323 M_\odot$ and $M_2 = 0.335 M_\odot$ derived from photometric measurements  \cite{2020ApJ...905L...7B}. The most relevant feature of ZTF J2243+5242 for the present analysis is that neither WD component is close to fill its Roche lobe, which allows a cleaner a simpler analysis of the binary dynamics.

We have shown that the dynamics of DWDs is largely affected by the UI for $\tilde{B} = \sqrt{(1-\alpha) \eta} B$ in the range $10$--$100$ MG. For large values of $\eta = 10^2$--$10^3$  (see Fig. \ref{fig:etamax}), the above implies that the binary dynamics might deviate from the pure GW-driven dynamics even for moderate values of the magnetic field strength $B\gtrsim 10^6$ G. Those fields are detectable by Zeeman splitting and features of the spectral absorption lines at optical and UV wavelengths (see, e.g., \cite{2015SSRv..191..111F}, for details). Magnetic fields near $\sim 1000$ MG shift the spectral lines at wavelengths far off their zero-field locations and show \textit{stationary} transitions (see, e.g., ZTF J1901+1458 in \cite{2021Natur.595...39C}). In the case of SDSS J0651+2844, ZTF J1539+5027, and ZTF J2243+5242, strong magnetic fields in the luminous components are ruled out by the absence of Zeeman splittings in the cores of the Balmer absorption lines. However, as we have shown above, the UI might still be present and affect the orbital dynamics for high values of $\eta$, leading to a high effective magnetic field $\tilde{B}$. Therefore, the measurement of the magnetic field strength of a high-magnetic WD in a close DWD via measured Zeeman splitting would become a compelling target for follow-up timing to test the theoretical framework presented in this work.


\section{Acknowledgments}
G.A.C. thanks financial support from the Coordena\c c\~ao de Aperfei\c coamento de Pessoal de N\'ivel Superior (CAPES) under the grants PDSE 88881.188302/2018--01 and PNPD 88887.368365/2019--00. J.G.C. is grateful for the support of CNPq (311758/2021--5) and FAPESP (2021/01089--1). The research of R.C.A. is supported by Conselho Nacional de Desenvolvimento Cient\'{i}fico e Tecnol\'{o}gico (CNPq), grant number 310448/2021-2, and Serrapilheira Institute grant number Serra-1708--15022. She also thanks the support by L'Oreal Brazil, with the partnership of ABC and UNESCO, Brazil. J.G.C. and R.C.A acknowledge financial support by ``Fen\^omenos Extremos do Universo" of Funda\c{c}\~ao Arauc\'aria. R.V.L. thanks financial support by the U.S. Department of Energy (DOE) under the grant DE-FG02--08ER41533, and Universidad de Los Andes, Colombia. M.M is grateful for the support of  CAPES, CNPq and  project INCT-FNA Proc. No. 464898/2014-5. J.F.R. thanks financial support from the Patrimonio Aut\'onomo - Fondo Nacional de Financiamiento para la Ciencia, la
Tecnolog\'ia y la Innovaci\'on Francisco Jos\'e de Caldas (MINCIENCIAS - COLOMBIA) under the grant No. 110685269447
RC-80740--465--2020, project 6955.

%





\bibliography{sample631}{}
\bibliographystyle{aasjournal}



\end{document}